\documentclass[%
 reprint,
 amsmath,amssymb,
 aps,
]{revtex4-2}

\usepackage{siunitx}
\usepackage{graphicx}
\usepackage{dcolumn}
\usepackage{bm}
\usepackage{booktabs}       
\usepackage{cancel}
\usepackage{ulem}

\expandafter\let\csname equation*\endcsname\relax
\expandafter\let\csname endequation*\endcsname\relax

\begin{document}


\title{Experimental Access to Observing Decay from Extremely Long-Lived Metastable Electronic States via Penning Trap Spectrometry}

\author{Bingsheng Tu} 
\email{bingshengtu@fudan.edu.cn}
\author{Ran Si}
\email{rsi@fudan.edu.cn}
\author{Yang Shen}
\author{Jiarong Wang}
\author{Baoren Wei}
\author{Chongyang Chen}
\author{Ke Yao}
\email{keyao@fudan.edu.cn}
\author{Yaming Zou}
\affiliation{%
 Shanghai EBIT laboratory, and Key Laboratory of Nuclear Physics and Ion-Beam Application (MOE), Institute of Modern Physics, Fudan University, Shanghai 200433, China
}%

\date{\today}

\begin{abstract}
Long-lived ionic quantum states known as metastable electronic states in highly-charged ions (HCIs) are of great interest in fundamental physics. Especially, it generates transitions with very narrow natural linewidth which is a promising candidate for use in the next generation HCI atomic clocks to reach an accuracy below $10^{-19}$. A recent experiment reported in [\textit{Nature},\textbf{581}(7806) 2020]~\cite{schussler2020}, used Penning trap mass spectrometry to measure the energy of an extremely long-lived metastable electronic state, thus opening doors to search for HCI clock transitions. Building upon prior research, this study introduces an experimental proposal with the goal of measuring lifetimes of the metastable states beyond seconds. Our approach employs a sequential pulse-and-phase measurement scheme, allowing for direct observations of the decay processes from metastable electronic states through single-ion mass spectrometry in a Penning trap. This measurement poses a significant challenge to conventional techniques like fluorescence detection. To demonstrate the effectiveness of this method, we conducted a comprehensive simulation under real experimental conditions, yielding promising results in a specific scenario. Two suitable candidates are proposed for testing this method, and the state-of-the-art MCDHF theory are employed for accurate energy levels and transition rate calculations. Some future prospects in the experimental determinations of a wide range of energy and lifetimes of long-lived metastable electronic states, probing hyperfine and magnetic quenching effects on high-order forbidden transitions and search for highly quality HCI clock transitions are discussed.

\end{abstract}

\maketitle

\section{\label{sec:level1}Introduction}

Atomic electron transitions between two electronic energy levels represent a fundamental quantum mechanical process within atoms and are responsible for most of the light observed in our daily lives. The rate of these transitions was first elucidated by Einstein's coefficients in 1916 and subsequently expressed through Fermi's golden rule within the framework of quantum electrodynamics (QED). According to selection rules, the decay of excited states through allowed (electronic dipole, E1) transitions, typically occurs within the nanosecond range. However, when transitions are forbidden, excited states can exhibit significantly longer lifetimes, rendering them metastable. These metastable electronic states hold substantial importance in modern clocks and frequency standards due to their high-quality factor transitions~\cite{Kozlov2018,Ludlo2015}. Furthermore, they provide an isolated quantum system for applications in quantum computing and quantum simulation~\cite{Blatt2012}, and allow for stringent tests of the building blocks of the standard model and fundamental theories, e.g. Lorentz symmetry~\cite{sanner2019} and general relativity~\cite{bothwell2022}, while also serving as a platform to explore new physics~\cite{Safronova2018}. \\
Highly charged ions (HCIs) stand out as a promising candidate for establishing the next generation of frequency standards. Within HCIs, the outer electrons are tightly bound, resulting in a remarkable insensitivity to external electromagnetic perturbations, thereby enabling us to exceed the current accuracy achieved in clocks composed of singly charged ions or neutral atoms. While inter-shell transitions within HCIs typically occur in the x-ray region, some intra-shell spectral lines, such as fine structure transitions, appear in the visible or ultraviolet (UV) range that frequency combs can, in principle, access~\cite{cingoz2012}. Extensive efforts have been devoted to both theoretical and experimental exploration to identify suitable clock transitions in HCIs~\cite{Safronova2014,Yu2016,Yu2018,Cheung2020,bekker2019,Liang2021}. As the first realization of this new class of clock, a team at Physikalisch-Technische Bundesanstalt (PTB) presented their groundbreaking experimental campaign on the $2s^22p$ $^2\mathrm{P}_{3/2} \rightarrow {^2\mathrm{P}_{1/2}}$ magnetic-dipole (M1) fine-structure transition of boron-like $\mathrm{Ar}^{13+}$ ~\cite{micke2020,king2022}. \\
The finite lifetime $\tau$ or natural linewidth of excited atomic metastable state plays a crucial role on the stability of frequency standard~\cite{Peik_2006}. Transitions with longer lifetimes, e.g. over seconds, are generally more favorable. In the last decade, experiments based on detecting the spontaneous decay of extremely long-lived electronic metastable states have already been presented by different research groups with magneto-optical traps (MOT)~\cite{Hodgman2009} and ion traps~\cite{Rosenband2007,Shao2016}. Later, Lange et al.~\cite{Lange2021} introduced a novel approach for determining the longest lifetime of the $^2F_{7/2}$ level in $\mathrm{Yb}^+$ by coherently driving the transition with a single trapped ion, completely bypassing any reliance on decay processes. These methods require the laser excitation with precisely known wavelengths and laser cooling of the trapped ions, which may not be straightforward to directly implement with HCIs. To date, fluoresces detection and population losses measurements of the excited states are two widely utilized techniques in ion traps~\cite{Guise2014,Brewer2018,Lapierre2006,Trabert2007,Beiersdorfer2016,Bao2023} and storage rings (see~\cite{Schippers2007,Trabert2010} and reference therein), for studying HCIs' lifetimes ranging from microseconds to hundreds of milliseconds. However, due to limitations in the detection efficiency of time-resolved photons or particles and the loss mechanisms associated with metastable states, which can arise from ion escape and ion collisions, measuring longer lifetimes extending beyond the seconds range presents a significant experimental challenge in HCIs.\\
Recently, an experimental group at the Max-Planck-Institut f\"ur Kernphysik (MPIK) performed a direct detection of metastable electronic state $4d^94f$ $^3\mathrm{H}_5$ of $^{187}\mathrm{Re}^{29+}$ by Penning trap mass spectrometry which opens a new door to search for high quality HCI clock transitions~\cite{schussler2020}. By employing an elaborated cyclotron frequency ratio (CFR) measurement, the energy of this metastable electronic state was determined with an accuracy below \SI{2}{\electronvolt} or \SI{500}{\tera\hertz} via a direct measure of mass difference of ions in the ground state and metastable electronic state. In this paper, we propose an expanded experimental method for observing spontaneous decay from extremely long-lived metastable electronic state with a single trapped ion. Instead of detecting decay photons, the motional frequency of the ion is continuously measured. Once the ion emits a photon that carries energy away, an equivalent mass change can be directly observed via the proposed sequential pulse-and-phase (Seq-PnP) method, and consequently the decay time is recorded. This approach enables precise measurements of both energy levels and natural linewidth of potential clock transitions in HCIs with lifetimes from seconds to days.\\
The transitions from metastable electronic states can be strongly impacted by an electromagnetic field generated from either the superconducting magnet or the nucleus. The latter effect is also known as hyperfine interaction. The magnetic and hyperfine induced transitions have been observed in some previous experiments~\cite{Trabert2007,Beiersdorfer2016}  via fluorescence detections. For extremely long-lived metastable electronic state of which the decay photon is barely detectable, the proposed technique also allows to probe hyperfine and magnetic quenching effect by measuring a reduction on its lifetime in Penning traps. \\
The rest of this paper is organized as follows. In section~\ref{sec:level2}, the measurement principle of penning trap spectrometry is briefly described and a simulation of measurement scheme based on a Seq-PnP method is presented on how the proposed technique works in the determination of energy and lifetimes of metastable electronic states. In section~\ref{sec:level4}, an accurate theoretical calculation using MCDHF method is implemented to give proper candidates of which the energy and lifetimes can be measured by the proposed method. In section~\ref{sec:level5}, we discuss the future prospects in measuring long-lived metastable electronic states and search for highly quality clock transitions. In section~\ref{sec:level6}, we conclude this proposal work. 

\section{\label{sec:level2}Measurement Principle}

\subsection{\label{sec:level21}Penning Trap Principle}

\begin{figure}[th!]
\centering
\resizebox{0.45\textwidth}{!}{
\includegraphics{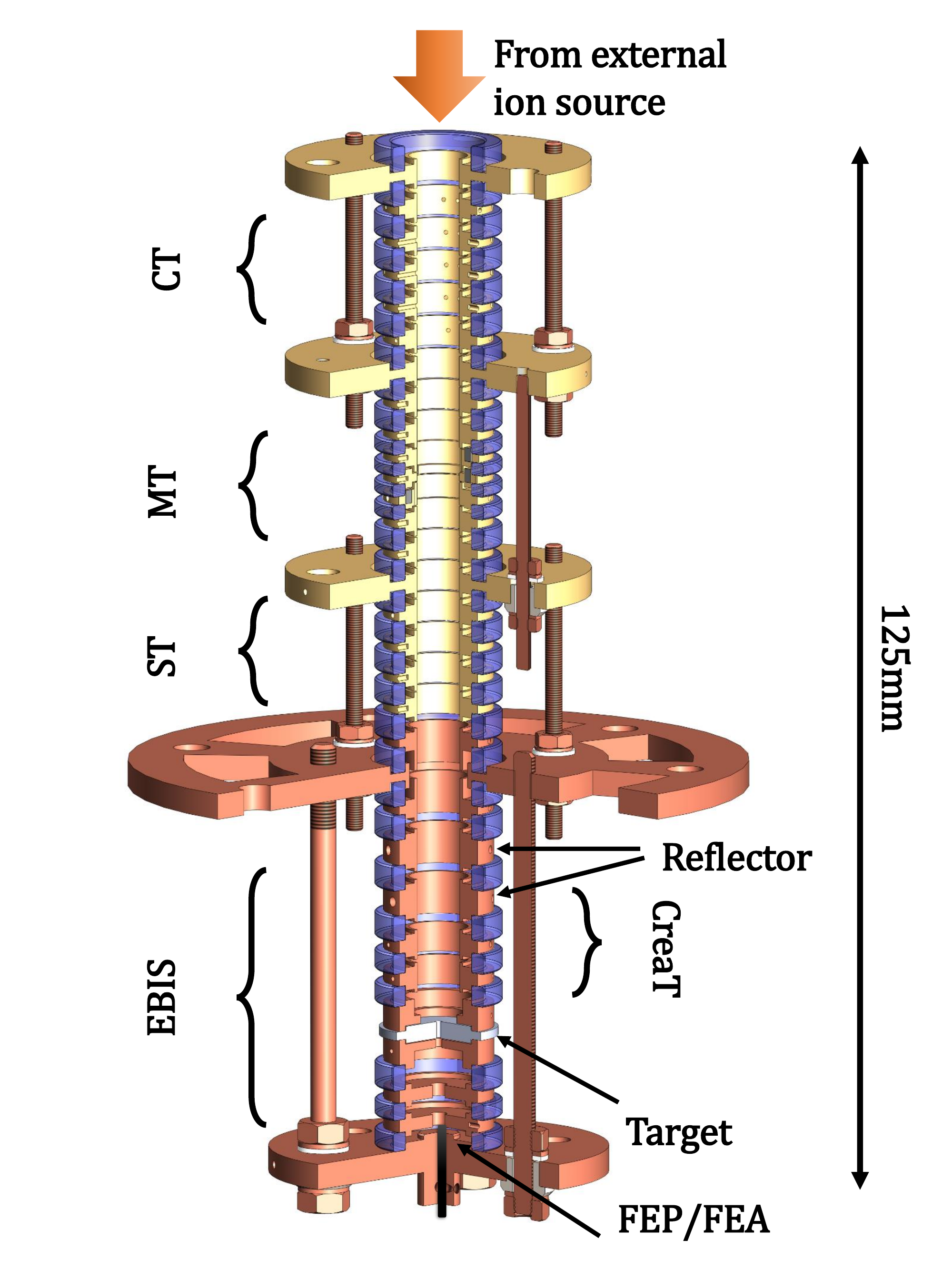}
}
\caption{Schematic drawing of the trap model. From top to bottom the capture trap (CT), measurement trap (MT), storage trap (ST) and in-trap EBIS which consists of a field emission point or array (FEP/FEA), a target, a creation trap (CreaT) and reflector electrodes. For details see text. }
\label{fig:1}
\end{figure}

The principle of Penning trap experiments has been well described in depth by other pioneers of Penning trap groups~\cite{sturm2019,Heisse2019,smorra2015,Fan2023,Fink2020,Goodwin2016,Jordan2019}. The experimental setup consists of a homogeneous magnetic field superimposed with an electrostatic quadrupole potential. The magnetic field provides a radial confinement of charged particles while the electrostatic potential prevents ions’ escape in axial direction. The resulting ion motion is a superposition of three harmonic oscillations: the radial modes split up into a fast modified cyclotron motion with frequency $\nu_+$ and slow magnetron motion with frequency $\nu_-$, and an axial harmonic oscillation with frequency $\nu_z$. In the combination of these eigen frequencies, the so-called invariance theorem $\nu_c= \sqrt{\left(\nu_+^2+\nu_z^2+\nu_-^2\right)}$~\cite{Brown1982} gives the free cyclotron frequency that is relevant to the charge to mass ratio by $\nu_c=\frac{Bq}{2\pi M}$. \\
The proposed lifetime measurement on metastable electronic states is planned to be performed at \textbf{Sh}anghai Penning \textbf{Trap} (SH-Trap) which is under construction. The trap tower is structured by using a cylindrical stack of electrodes like the predecessor experiments~\cite{sturm2019,Heisse2019} (Fig.~\ref{fig:1}), located in the worm bore of the magnet and cooled down to \SI{4}{\kelvin} with the help of an ultra-low-vibration cryostat. The electrostatic potential produced by this cylindrical trap design could deviate from the ideal quadrupole potential which is called electric field imperfections. The general potential in the center of the measurement trap (MT) can be written as a series expansion:

\begin{equation}
U\left(r,\theta,z\right) = \frac{U_0}{2}\sum_{n=0}^{\infty} C_n\left(\frac{r}{d_\mathrm{char}}\right)^n P_n\left(cos\theta\right),
\label{eq:2}
\end{equation}

\noindent where, $C_n$ represents a dimensionless expansion coefficient and $P_n(cos\theta)$ is the Legendre polynomials. To achieve the desired quadrupole potential with only a non-zero $C_2$, pairs of correction electrodes can be stacked on both sides of the (center) ring electrode. However, due to the limited space and manufacture imperfections, typically only the main high order coefficients $C_4$ and $C_6$ are optimized. With a 7-mm-diameter five-electrode trap design the electric potential coefficients $|C_2|=0.55$, $|C_4|<10^{-5}$ and $|C_6|<10^{-2}$ can be achieved as a similar design being already described in Mainz~\cite{Heisse2019}. \\

\noindent A superconducting solenoid magnet will be manufactured to produce a homogeneous magnetic field upto \SI{7}{\tesla}. The main magnetic field imperfections can be expressed by a linear magnetic field gradient $B_1$ as well as the quadratic dependency $B_2$. Both the electric and magnetic field imperfections can cause undesirable frequency shifts on eigen motions, which contributes systematic uncertainties in many precision measurements at Penning trap facilities~\cite{ketter2014}. Therefore, more care should be taken to evaluate the effects on the energy and lifetime measurements of metastable electronic states proposed in this work, for details see chapters below.

\subsection{\label{sec:level22}Detection method}

Highly charged ions can be produced by an in-trap electron beam ion source (EBIS) on the bottom side of the trap tower, see Fig.~\ref{fig:1}. The electron beam which emits from a field emission point or array (FEP/FEA) can be reflected by the reflector electrodes and then hit on a target to sputter the atoms out. Those atoms will continuously collide with electrons such that they become highly charged and trapped in the Creation Trap. This procedure is successfully tested in Mainz~\cite{kohler2016,rau2020}. However, the charge state is limited by the maximum bias on the electrodes due to the compact structure of the trap and feedthroughs. For ions of higher charge states, they could be produced in an external ion source. A room temperature permanent magnet electron beam ion trap called CUBIT has been developed for HCIs extraction, for details see~\cite{Li2021,WANG2022,HE2022}. After the HCIs inject into the capture trap with a kinetic energy of about a few hundreds eV, the voltage on the electrodes can be quickly pulsed in order to capture the incoming ions. With the help of cryogenic pumping and a dedicated cryo-valve, most residual gas in the trap chamber is frozen resulting in a very good vacuum of $10^{-16}$ and thus HCIs can stay long, e.g. a few months for $\mathrm{Ar}^{13+}$ for measurements (see the ALPHATRAP experiment~\cite{sturm2019}). \\
Once the ions are captured in CT or created in CreaT, they will be transported to MT. Then, a single HCI will be produced after a dedicated purification procedure (see~\cite{sturm2019,Heisse2019}). The ion's oscillation frequencies can be detected by a so-called image current detection method. The image current induced by ion's oscillation on one of the trap electrodes can be picked up by a superconducting tank circuit (resonator) which is tuned close to ion's oscillation frequencies. In this way, the weak current signal, across the equivalent parallel resistance $R_p$, is converted into a detectable voltage signal . Before undergoing the Fast Fourier Transformation (FFT), the signal must first be amplified using low-noise cryogenic and room-temperature amplifiers and then mixed down with a local oscillator to a frequency of approximately \SI{10}{\kilo\hertz}. Fig.~2(a) shows the typical dip signal (simulated) of the axial motion of a single ion represented by $\mathrm{Fe}^{8+}$ when it is in the thermal equilibrium with the detector circuit. While the ion is axially excited, the peak signal instead of the dip signal appears upon the resonator noise, see Fig.~2(b). Two radial mode frequencies can be measured by coupling either modified cyclotron motion or magnetron motion to axial motion via a quadrupole excitation with radio frequency at $\nu_{\mathrm{rf}} = \nu_+ - \nu_z$ or $\nu_{\mathrm{rf}} = \nu_- + \nu_z$. During the coupling, the axial and radial motion start to exchange energy undergoing a rabi-type oscillation. In this case, two radial motions are cooled and eventually thermalized with the tank circuit as well and the axial dip splits into two dips which is so-called double dip spectrum (see Fig.~2(c)). By measuring the left and right dips, the modified cyclotron frequency and magnetron frequency can be obtained. \\

\begin{figure*}[th!]
\centering
\resizebox{1\textwidth}{!}{
\includegraphics{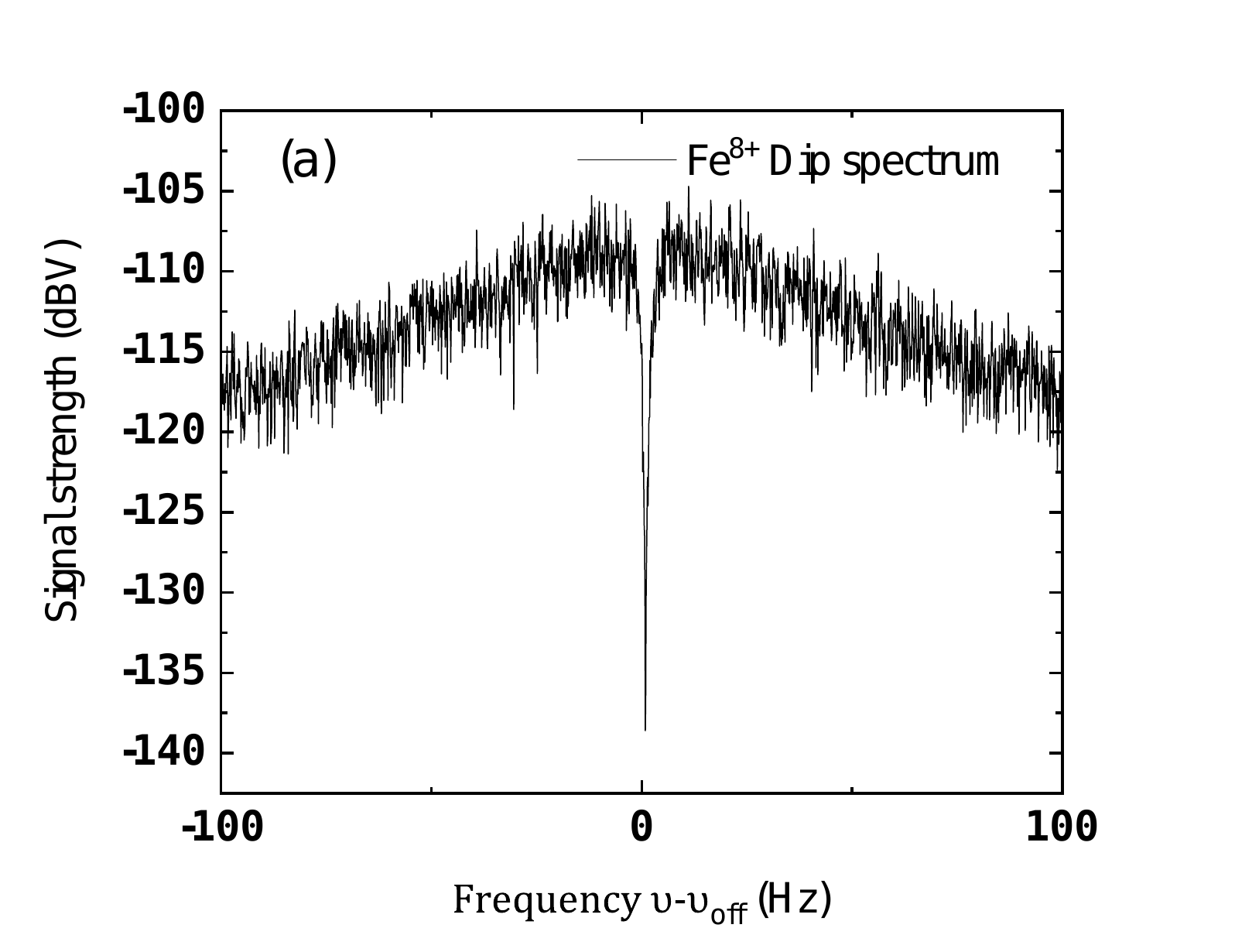}%
\includegraphics{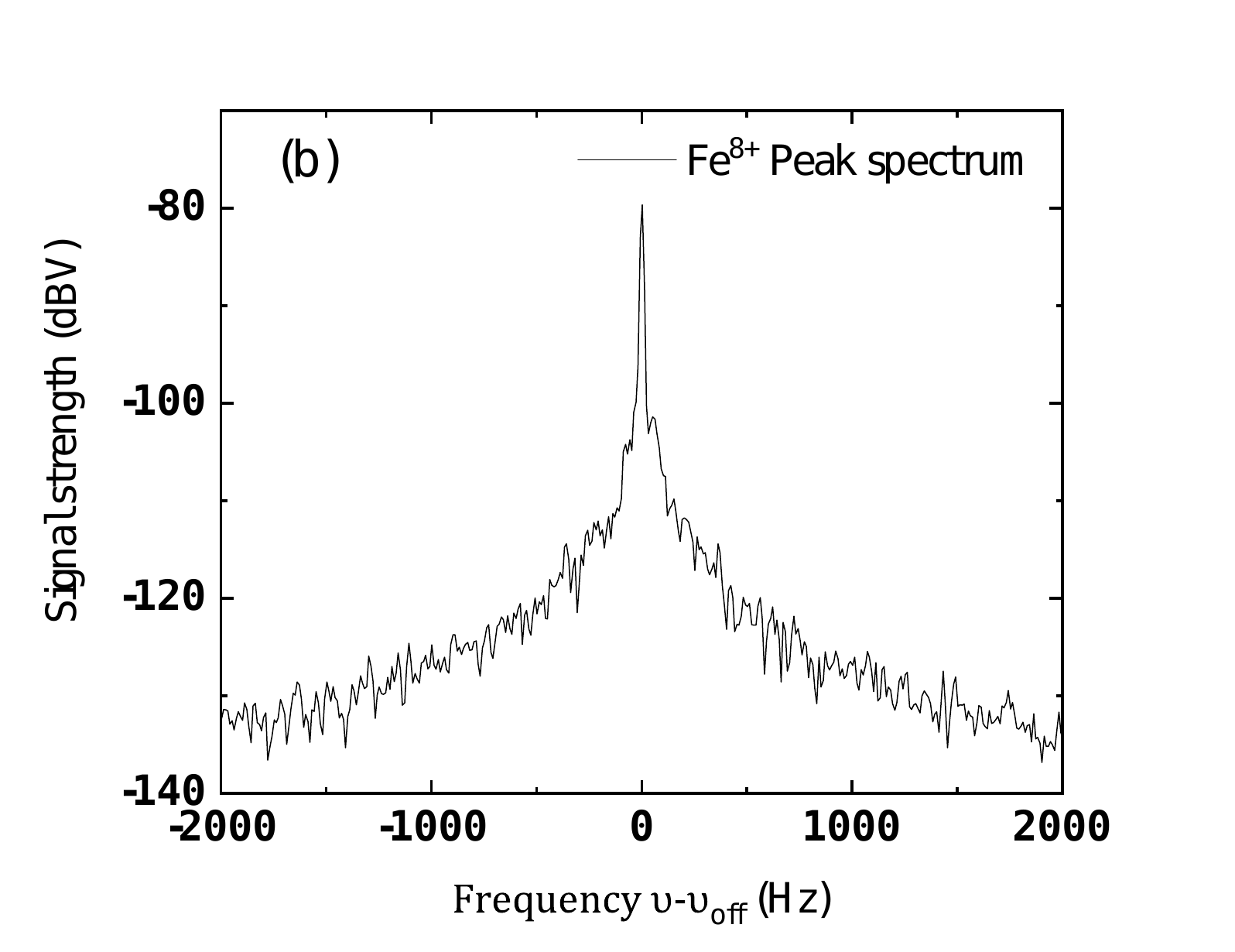}%
\includegraphics{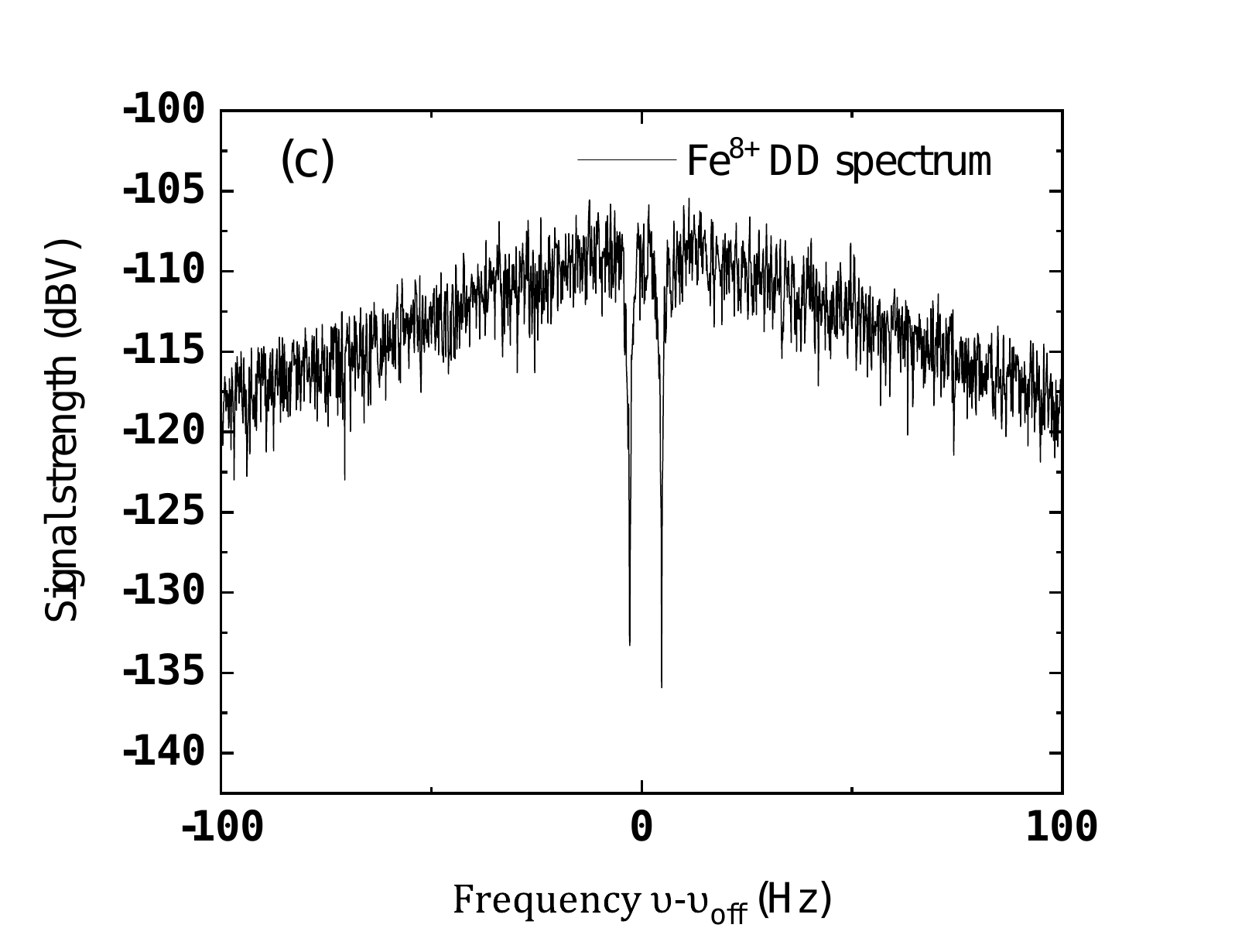}%
}
\caption{The simulated FFT spectra of a single $\mathrm{Fe}^{8+}$ ion detected by a tank circuit with $Q$ value of about 5000. (a) the axial dip signal, (b) the axial peak signal and (c) double dip signal with $\nu_{\mathrm{rf}} = \nu_- + \nu_z $ coupling. The dip widths are related to charge to mass ratio, $Q$ value and trap geometry. }
\label{fig:2}
\end{figure*}

\subsection{\label{sec:level3}Measurement scheme}

\begin{figure*}[th!]
\centering
\resizebox{0.8\textwidth}{!}{
\includegraphics{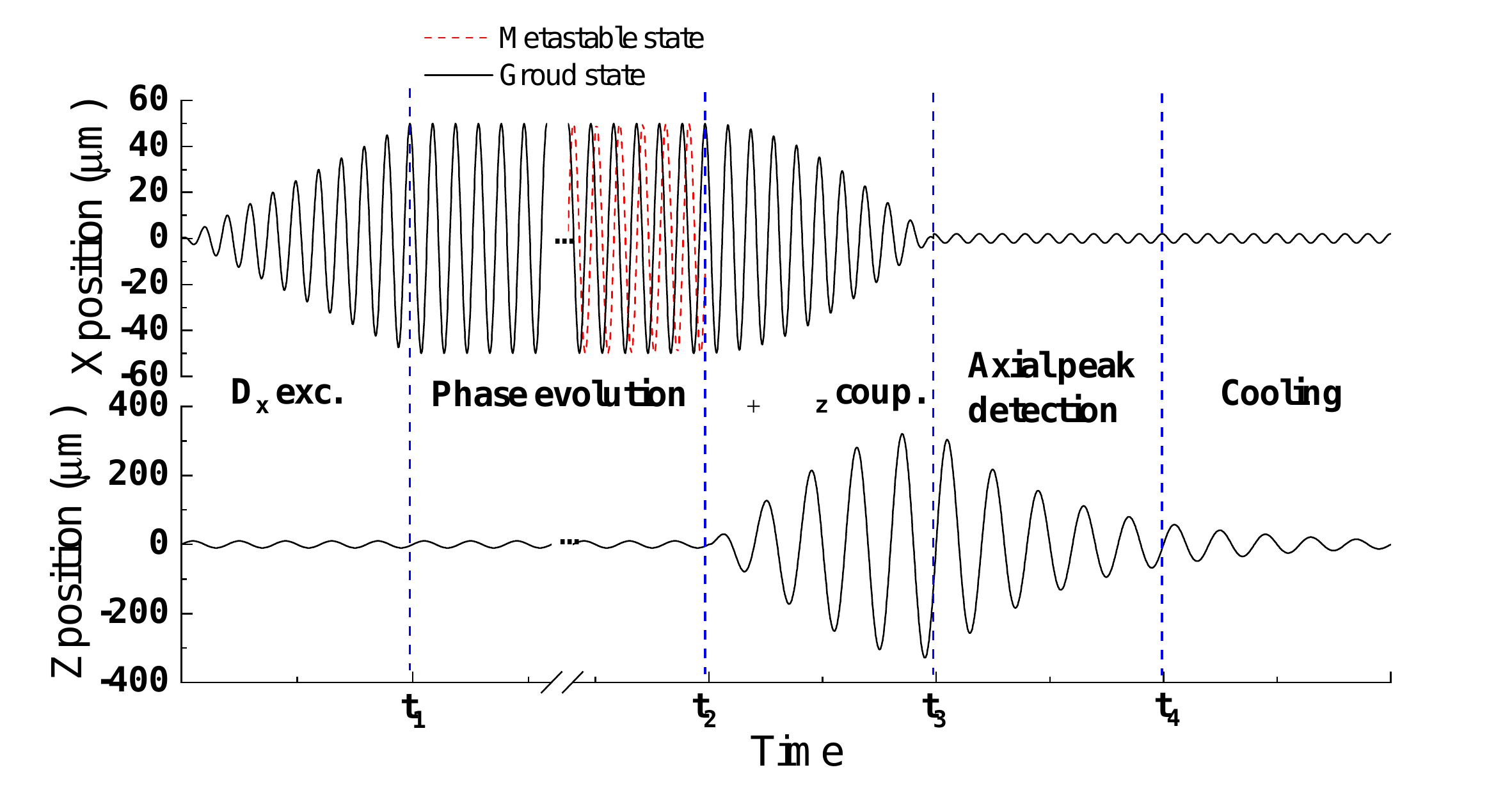}
}%
\caption{Diagram of ion motions for the measurement scheme, see radial motion in X position and axial motion in Z position (the frequencies are rescaled in plot for better visibility). The modified cyclotron motions of metastable electronic state (red dash line) and ground state (black line) show a phase difference during the evolution time.}
\label{fig:3}
\end{figure*}

One typical double dip spectrum enables an accurate determination of the cyclotron frequency at a few $10^{-9}$ level. It is still hard to distinguish some long-lived metastable electronic states, for instance $3s^23p^53d$ $^3\mathrm{F}_4$ state of $^{56}\mathrm{Fe}^{8+}$ of which the energy is about \SI{52.79}{\electronvolt}. In addition, recording FFT spectrum takes around 100 seconds, which is not suitable for measuring short lifetimes. For a better frequency resolution, a pulse and phase sensitive method ~\cite{Heisse2019,Sturm2011} can be used, and a relative cyclotron phase jitter of below $2\times10^{-10}$ from pulse to pulse can be achieved. To observe decay from metastable electronic states, instead of taking tens of minutes to implement normal PnP method introduced in~\cite{schussler2020}, we propose a sequential PnP method continuously measuring the variance of modified cyclotron frequency. Because of $\nu_+ \approx \nu_c$ , if a phase jump on the modified cyclotron frequency is observed, it indicates that the ion mass has changed, or in other words, the decay occurs. \\
Once a single cold ion is prepared in the trap, an electron beam emitted from the FEA or FEP, with precise voltage adjustment, is utilized to collide with the ion, leading to its excitation to the long-lived metastable electronic state known as electron impact excitation (EIE). After that, the $^{56}\mathrm{Fe}^{8+}$ ion is quickly thermalized with the tank circuit and then the measurement scheme (see Fig.~\ref{fig:3}) starts including several steps in one run: (1) Excite the modified cyclotron motion with a starting phase $\varphi_0$. (2) Wait for cyclotron phase evolution with $t_\mathrm{evol}$. The phase ends with $\varphi(t_\mathrm{evol})=\varphi_0+360^{\circ}\nu_+ t_\mathrm{evol}$. For a well-defined $\varphi_0$ and fixed $t_\mathrm{evol}$, the ending phase only depends on $\nu_+$ corresponding to which electronic state the ion stays. (3) Implement a red sideband $\nu_\mathrm{rf}=\nu_+ - \nu_z$ quadrupole pi-pulse coupling between modified cyclotron motion and axial motion to transfer the cyclotron amplitude and phase to axial motion for readout. (4) Measure the axial peak (axially hot ion) signal to obtain the transferred cyclotron phase. (5) Cool axial motion and start next cycle. \\
\noindent According to the envisaged electromagnetic field settings in SH-Trap ($B = $ \SI{7}{\tesla} and $U_0 = $ \SI{-5.7}{\volt}) and the mass value of $^{56}\mathrm{Fe}^{8+}$ from AME 2020~\cite{Huang_2021}, the frequencies are $\nu_+ \sim$ \SI{15}{\mega\hertz}, $\nu_z \sim$ \SI{350}{\kilo\hertz} and $\nu_- \sim$ \SI{4}{\kilo\hertz}. In the first step, the ion is excited to \SI{50}{\micro\meter} in cyclotron radius. Then it freely evolutes for $t_\mathrm{evol}$  $\sim$ \SI{10}{\second} giving rise to a phase difference $\Delta \varphi \sim 54$ degrees if ion stays in difference electronic states. With a careful control of the quadrupole pi pulse, cyclotron energy can be fully transferred to axial mode. Finally, the axial motion with an amplitude of about \SI{330}{\micro\meter} is detected and then cooled by the tank circuit. By sequentially implementing this measurement scheme, the modified cyclotron phase for a fixed $t_\mathrm{evol}$ is continuously recorded such that the variance of the cyclotron frequency can be observed immediately once the ion mass changes due to the decay. \\
To perform CFR measurement in the same measurement scheme, we refer to an alternative solve of mass ratio $R$ ~\cite{Cornell1992}:
\begin{equation}
R = \frac{\nu_{z1}^2}{2\nu_{c1}^2}+(1-\frac{\nu_{z1}^2}{2\nu_{c1}^2})\left(1+\frac{\Delta \nu_+(\Delta \nu_+ - 2 \nu_{+1})}{(\nu_{c1}-\nu_{z1}^2 /2 \nu_{c1})^2}\right)^{1/2}
\label{eq:cfr}
\end{equation}
where, we define subscript $1$ the ground state. $\Delta \nu_+ = \nu_{+,g}~-~\nu_{+,m} = \frac{-\Delta \varphi}{360^{\circ}t_\mathrm{evol}}$ is the modified cyclotron frequency difference, which is measured from the variance of the cyclotron phase. Just after the decay, a standard PnP measurement with short evolution time can be implemented to determine $\nu_{+1}$ or $\nu_{+,g}$. Dip and double dip spectra can be recorded to measure $\nu_{z1}$ and  $\nu_{-1}$, thereby providing the necessary information to determine $\nu_{c1}$. Because $\Delta \nu_+$ and $\nu_{+1}$ are measured back-to-back, the effect from magnetic field instability is significantly suppressed, allowing for an even higher accuracy in the comparison to the standard CFR measurement involving ions transport. 

\subsection{\label{sec:level31}Numerical implementation}
Here, we present a numerical implementation for the proposed Seq-PnP measurement scheme by solving the equations of ion motion in the external electromagnetic field from first principles. The numerical model consists of a single ion oscillating in a static electromagnetic trapping field, a tank circuit for image current detection and external electric pulses for ion motion operation. The differential equations of this model can be expressed by: 

\begin{equation}
\begin{aligned}
&\ddot{x} = \frac{q}{M}\left(-\frac{\partial U}{\partial x} + B_z \dot{y} - B_y \dot{z} + E_x\left(x,y,z \right) \right) \\
&\ddot{y} = \frac{q}{M}\left(-\frac{\partial U}{\partial y} - B_z \dot{x} + B_x \dot{z} + E_y\left(x,y,z \right) \right)\\
&\ddot{z} = \frac{q}{M}\left(-\frac{\partial U}{\partial z} + B_y \dot{x} - B_x \dot{y} + E_z\left(x,y,z \right) + \frac{L\dot{I_L}}{D_\mathrm{eff}} \right)\\
&\ddot{I_L} = \frac{1}{LC_p}\left( -\frac{q}{D_\mathrm{eff}}\dot{z} - I_L -I_\mathrm{noise} -\frac{L}{R_p}\dot{I_L} \right).
\end{aligned}
\label{eq:4}
\end{equation}

\noindent Here $D_\mathrm{eff}$ is the effective distance of the electrode. $U$ and $B$ are the electric potential and magnetic field strength, respectively, and the main field imperfection coefficients $C_4$, $C_6$ and $B_2$ are included, while the effect from the odd terms such as $C_3$, $C_5$ can be neglected due to the trap symmetry. On top of that, a voltage noise and magnetic field fluctuation are included by adding a random walk noise upon the main field. $E_{x,y,z}(x,y,z)$ denotes the external electric excitation. To take into account the interaction between a single ion and a superconducting tank circuit, a treatment of image current detection has been described in ref~\cite{Will_2022}. $C_p$ is the equivalent parallel capacitance of the detection circuit. $I_L$ is the current  and $I_\mathrm{noise} = N(1,0) \sqrt{\frac{2k_B T_{LC} L}{R \triangle t}}$  is the thermal Johnson noise current through the inductance of the tank circuit, where $T_{LC}$ is the environmental temperature and $N(1,0)$ is a number sampling from a Gaussian distribution. Selecting an appropriate simulation time step is crucial; for instance,  $\triangle t=$ \SI{0.2}{\nano\second} is suitable for ions such as $^{56}\mathrm{Fe}^{8+}$ with a cyclotron frequency $\nu_+ \sim $ \SI{15}{\mega\hertz}. This setting ensures that the finite time step has negligible impact on the accuracy of the numerical solution, while maintaining a manageable computational time.\\
The set of the differential equations is solved by employing a fourth-order Runge-Kutta method (RK4) which is a widely-used numerical integration algorithms for finding solutions in initial value problems. For an efficient calculation, the RK4 solver is written by C++ language and compiled into a package which is then called by the main script written by Python. In the main script, all the variables in the simulation are initialized and then sent to RK4 solver for calculation. After the differential equations are solved, the time dependent data is down mixed by a sinusoidal signal with a frequency at $\nu_z-$ \SI{11.8}{\kilo\hertz} and then sampled with a rate of \SI{200}{\kilo\hertz} for FFT spectrum in order to determine the motional frequencies and phases. 

\subsection{\label{sec:level32}Simulation results and benefits}

\begin{figure}[th!]
\centering
\resizebox{0.5\textwidth}{!}{
\includegraphics{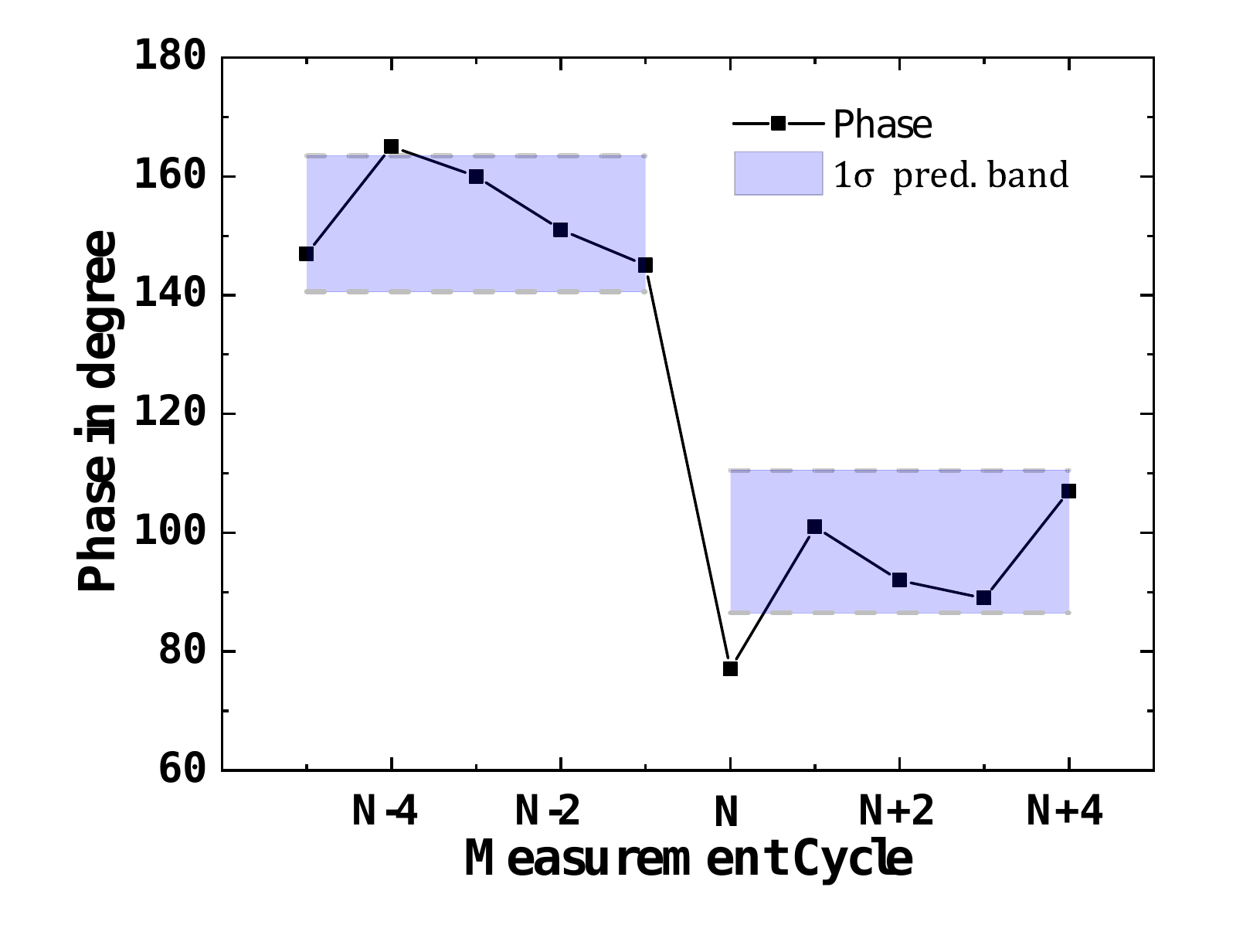}
}%
\caption{Simulation results of Seq-PnP measurement of metastable electronic state $3s^23p^53d$ $^3\mathrm{F}_4$ of $^{56}\mathrm{Fe}^{8+}$, assuming a decay taking place in cycle N. The light blue areas indicate the $1\sigma$ prediction band of the simulated modified cyclotron phases after a free evolution time of \SI{10}{\second}, with a total number of 60 simulation runs for each electronic state.}
\label{fig:4}
\end{figure}

The simulation results of the lifetime determination of metastable electronic state $^3\mathrm{F}_4$ of $^{56}\mathrm{Fe}^{8+}$ via a Seq-PnP measurement is shown in Fig.~\ref{fig:4}. The equivalent mass value of this excited state is converted from its excitation energy. The modified cyclotron phase evolution time is 10 seconds which causes a phase difference of about 54 degrees between the metastable and ground state. Once the decay takes place at measurement cycle N, a phase jump can be resolved and the simulated phase jitter is around $11\sim 12$ degrees equivalent to about $2\times 10^{-10}$ relative fluctuations. \\
\noindent In order to verify the feasibility of this measurement technique, the main systematic uncertainties are taken into account, conducting a simulation close to real experimental conditions. \\
(1) The thermal motion can affect the modified cyclotron phase from several aspects. Firstly, the imprinted phase from the radial dipole excitation can be disturbed by the random phase associated with the thermal motion. Secondly, after this excitation the cyclotron amplitude can be written as $\overrightarrow{r_+} = \overrightarrow{r_\mathrm{exc}} + \delta \overrightarrow{r}$, where $\delta \overrightarrow{r}$ denotes a jitter attributed to the thermal motion. Then, this amplitude jitter causes the frequency jitter due to the field imperfections and special relativity. In the present model, the ion is thermalized with the detection circuit to about \SI{4.2}{\kelvin}, and thus the initialized ion motional amplitudes and phases are drawn from a Boltzmann distribution at this temperature. The field imperfections are conservatively set, with values $C_4 = 10^{-5}$, $C_6 = 10^{-2}$ and $B_2 =$ $0.2$ $\mathrm{T}/\mathrm{m}^2$.  The special relativity is not included in the simulation but its effect is calculated to a relative uncertainty at $2 \times 10^{-11}$ level. \\
\noindent (2) During the free evolution time, any fluctuation on the electromagnetic field can disturb the modified cyclotron phase. The simulation model has already included a random walk noise based on an achievable noise feature of voltage stability of $4\times 10^{-8}$ (UM1-14 power supply by Stahl Electronics) and B field stability of $2\times 10^{-10}$ in a time span of 10 seconds. The effect from a magnetic field drift typically $<10^{-9}$ per hour can cause a frequency drift only at $10^{-12}$ level in one measurement run and therefore it is not included in the simulation model for simplicity. \\
\noindent (3) The technical phase jitter during the axial peak signal detection is also simulated in our model. The signal-to-noise ratio (SNR) of the axial peak spectrum Fig. 2(b) which is synthesized by employing a FFT analysis leads to a major phase jitter of a few degrees. The axial amplitude can be optimized to reduce the SNR jitter until the phase jitter caused by the field imperfections or special relativity becomes sizeable. In the present model, the axial amplitude is about \SI{330}{\um} after a quadrupole pi-pulse coupling corresponding to a SNR of about 25 dB and the phase jitter from the field imperfection effect is still small within a readout time of 0.1 second which is about three times of the cooling time constant of the ion. In addition, the impact on the phase determination from the axial thermal motion is negligible due to the relatively small amplitude of about \SI{10}{\um}. The axial frequency fluctuates because of the voltage supply stability, which in principle leads to another readout phase jitter, but this effect is also small in the comparison with other sources of jitter.\\
\noindent (4) Some other negligible effects we do not include in our model: i) The recoil effect from the decay photon to the heavy ion like $^{56}\mathrm{Fe}^{8+}$. ii) The quantum state jump in cyclotron motion due to the electric field noise. iii) The image charge effect is not included since it cancels for the ions with same charges and nearly identical masses.\\
As a result, we have run the simulation model including all the systematics discussed above for a short evolution time of \SI{5}{\milli\second} and a long evolution time of \SI{10}{\second}. The simulated short-term phase jitter about 5 degrees comes from the axial peak signal readout, while the long-term phase jitter about $11\sim 12$ degrees (also see the prediction band in Fig.~\ref{fig:4}) is mainly attributed to the magnetic field fluctuation. From the simulation verification, the Seq-PnP method enables to search for decay of long-lived metastable electronic states like $^3\mathrm{F}_4$ of $^{56}\mathrm{Fe}^{8+}$ with equivalent mass difference at $10^{-9}$ level in 10 seconds. What's more, in comparison to other lifetime measurements which involves the detection of decay from a multitude of ions, the accuracy of this single-particle measurement technique is primarily constrained by statistical factors, giving opportunities to test the atomic structure theory and cross-verify existing lifetime measurements with a high degree of accuracy.


\section{\label{sec:level4}Atomic Structure Calculations}

\subsection{\label{sec:level41}MCDHF method}
For the theoretical studies on energy and lifetimes of long-lived metastable electronic states, a state-of-the-art theoretical method called the multiconfiguration Dirac–Hartree–Fock (MCDHF) method~\cite{Fischer_2016} has been employed. In the MCDHF method, as implemented in the GRASP package~\cite{JONSSON2013,FROESEFISCHER2019}, the atomic state function (ASF), is expanded in antisymmetrized and coupled configuration state functions (CSFs). 
\begin{equation}
|\Gamma J \rangle=\sum_{\gamma} c_{\gamma}|\gamma J\rangle,
\label{eq:5}
\end{equation}
\noindent where $\gamma$ represent all the coupling tree quantum numbers needed to uniquely define the CSF. The CSFs are four component spin-angular coupled, antisymmetric products of one-electron Dirac orbitals.\\
\noindent The radial parts of the Dirac orbitals together with the mixing coefficients are obtained in a relativistic self-consistent field (RSCF) procedure in the extended optimal level (EOL) scheme~\cite{DYALL1989}. The angular integrations needed for the construction of the energy functional are based on the second quantization method in the coupled tensorial form~\cite{Gaigalas_1997,GAIGALAS2001}. The transverse photon (Breit) interaction and the leading quantum electrodynamic (QED) corrections (vacuum polarization and self-energy) can be accounted for in subsequent relativistic configuration interaction (RCI) calculations. In the RCI calculations, the Dirac orbitals from the RSCF step are fixed and only the mixing coefficients of the CSFs are determined by diagonalizing the Hamiltonian matrix. \\
The transition rate $A$ (s$^{-1}$) between two states $|\Gamma'J'\rangle$ and $|\Gamma J\rangle$ are expressed in terms of reduced matrix elements of the relevant transition operators~\cite{cowan1981}, where the reduced transition matrix element is the square root of the line strength $S$ multiplied with a factor. For the electric dipole (E1) transitions,

\begin{equation}
A(\Gamma' J' \rightarrow \Gamma J)=\frac{2.0261\times10^{18}}{(2J'+1)\lambda^3}|\langle \Gamma J|| \bm{P}^{(1)} || \Gamma' J' \rangle|^2,
\label{eq:7}
\end{equation}

\noindent where, $\lambda$ is the transition wavelength in \AA. In isotopes with non-zero nuclear spin, the interaction between the nuclear electric and magnetic multipole moments and the electrons couples the nuclear spin $I$ and the electronic angular momentum $J$ to a new total angular momentum $F$ and splits each fine structure level into several hyperfine levels. The corresponding Hamiltonian may be represented as a multipole expansion:

\begin{equation}
H_{\mathrm{hfs}}=\sum_{k\geq1}\bm{T}^{(k)}\cdot \bm{M}^{(k)},
\label{eq:8}
\end{equation}

\noindent where $\bm{T}^{(k)}$ and $\bm{M}^{(k)}$ are spherical tensor operators of rank $k$, operating on the electronic and nuclear parts of the wave function, respectively. The hyperfine interaction will introduce a mixing between states with different $J$ quantum numbers, and can open new decay channels for some transitions. The wavefunction of the atomic system can be written as an expansion: 

\begin{equation}
| \tilde{\Gamma} IF \rangle= \sum_{\Gamma J} d_{\Gamma J} | \Gamma IJF \rangle.
\label{eq:9}
\end{equation}

\noindent The E1 transition rate between two hyperfine states  $|\tilde{\Gamma'} IF'\rangle$ and $|\tilde{\Gamma} IF\rangle$ is given by

\begin{equation}
\begin{aligned}
&A(\tilde{\Gamma'} IF' \rightarrow \tilde{\Gamma} IF)\\
& = \frac{2.0261\times10^{18}\times(2F+1)}{\lambda^3} \bigg| \sum_{\Gamma J} \sum_{\Gamma' J'}d_{\Gamma J} d_{\Gamma' J'}\\
&\times  (-1)^{I+J'+F+1} \times \left\{ {\begin{array}{*{20}{c}}
J & F & I\\
F' & J' &1
\end{array}} \right\}
\langle \Gamma J|| \bm{P}^{(1)} || \Gamma' J' \rangle \bigg|^2.
\end{aligned}
\label{eq:10}
\end{equation}

\noindent In presence of the magnetic field, for isotopes with zero nuclear spin, only $M_J$ is a good quantum number, we can represent the wavefunction of the atomic system by 

\begin{equation}
| \tilde{\Gamma} M_J \rangle= \sum_{\Gamma J} d_{\Gamma J} | \Gamma JM_J \rangle.
\label{eq:12}
\end{equation}

\noindent The E1 transition rate between two states  $|\tilde{\Gamma}' M_J'\rangle$ and $|\tilde{\Gamma} M_J\rangle$ is given by

\begin{equation}
\begin{aligned}
& A(\tilde{\Gamma}' M_J' \rightarrow \tilde{\Gamma} M_J) \\
 = & \frac{2.0261\times10^{18}}{\lambda^3} \sum_q \bigg| \sum_{\Gamma J} \sum_{\Gamma' J'} (-1)^{J-M_J} d_{ \Gamma J} d_{\Gamma' J'}\\
& \times
\left( {\begin{array}{*{20}{c}}
J & 1 & J'\\
-M_J & q &M_J'
\end{array}} \right)
\langle \Gamma J|| \bm{P}^{(1)} || \Gamma' J' \rangle \bigg|^2.
\end{aligned}
\label{eq:13}
\end{equation}

\noindent For isotopes with non-zero nuclear spin, the wavefunction of the atomic system can be represented as, 

\begin{equation}
| \tilde{\Gamma} IM_F \rangle= \sum_{\Gamma JF} d_{\Gamma JF} | \Gamma IJFM_F \rangle.
\label{eq:14}
\end{equation}

\noindent The transition rate for an E1 transition between $|\tilde{\Gamma}' I M_F' \rangle$ and $|\tilde{\Gamma} I M_F \rangle$ is given by
\begin{equation}
\begin{aligned}
& A(\tilde{\Gamma}' IM_F' \rightarrow \tilde{\Gamma} IM_F) \\
 = & \frac{2.0261\times10^{18}}{\lambda^3} \sum_q \bigg| \sum_{\Gamma JF} \sum_{\Gamma' J'F'} (-1)^{F-M_F} d_{ \Gamma JF} d_{\Gamma' J'F'}\\
& \times \sqrt{(2F+1)(2F'+1)}
\left( {\begin{array}{*{20}{c}}
F & 1 & F'\\
-M_F & q &M_F'
\end{array}} \right)
(-1)^{I+J'+F+1}\\
& \times \left\{ {\begin{array}{*{20}{c}}
J & F & I\\
F' & J' &1
\end{array}} \right\}
\langle \Gamma J|| \bm{P}^{(1)} || \Gamma' J' \rangle \bigg|^2.
\end{aligned}
\label{eq:15}
\end{equation}

\noindent By diagonalizing the interaction matrix, hyperfine structure substates and Zeeman energy splittings are obtained together with the expansion coefficients of the basis functions, the transition rates can thus be computed. The other multiplolar transition rates between fine structure/hyperfine structure/Zeeman levels are not present here, and can be obtained
in a similar way. 

\subsection{\label{sec:level42}Calculations and Results}
\begin{figure}[th!]
\centering
\resizebox{0.5\textwidth}{!}{
\includegraphics{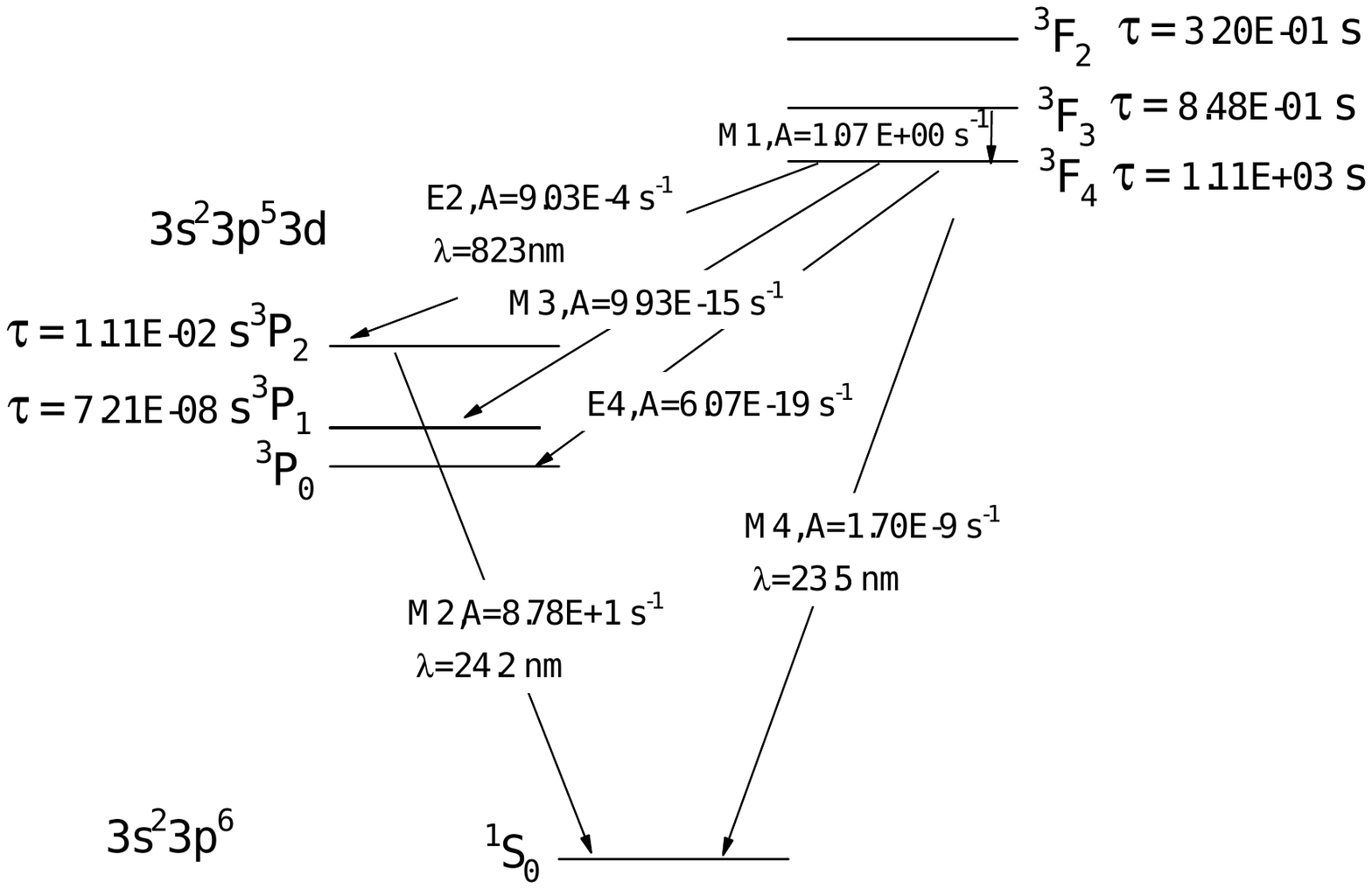}
}
\caption{Low-lying energy levels of $^{56}\mathrm{Fe}^{8+}$ based on MCDHF calculations. The decay rates from long-lived metastable electronic states $^3$F$_4$ as well as its lifetime has been shown. }
\label{figFe}
\end{figure}

\begin{figure}[th!]
\centering
\resizebox{0.5\textwidth}{!}{
\includegraphics{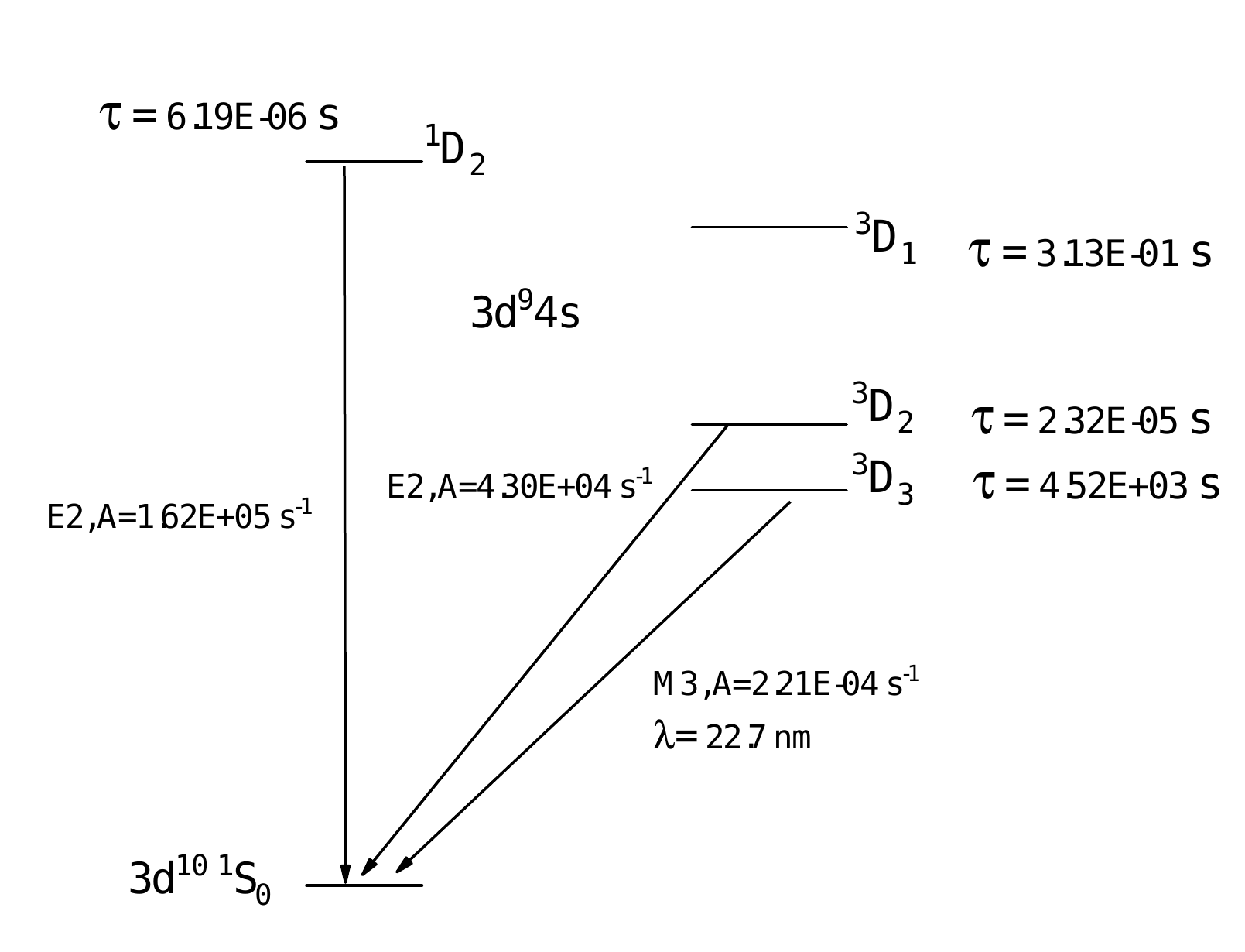}
}
\caption{Low-lying energy levels of $^{76,77}$Se$^{6+}$ based on MCDHF calculations. The decay rate from long-lived metastable electronic states $^3$D$_3$ as well as its lifetime has been shown. }
\label{figSe}
\end{figure}

Our calculations are based on the restricted active space (RAS) method, where the CSF expansions are obtained by allowing single and double (SD) substitutions from selected reference configurations to given orbitals to an active set (AS)~\cite{Jeppe1988}.
The active set increases by increasing the number of layers, specifically, a set of virtual orbitals specified by its principal quantum number. We sort the electron correlation effects into three types: i) substitutions only from the outermost valence subshells, valence-valence (VV) correlation is included; ii) at the most one substitution from the core subshell, core-valence (CV) correlation is accounted for; iii) double substitutions from the core subshells are allowed, core-core (CC) correlation is included.\\
The low-lying levels of interest in this work for $^{56}\mathrm{Fe}^{8+}$ and $^{76,77}$Se$^{6+}$ are shown in Fig.~\ref{figFe} and \ref{figSe}. As a starting point RSCF calculations were done for $3s^22p^6$, $3s^23p^53d$ configurations of $^{56}\mathrm{Fe}^{8+}$, and for the $3d^{10}$, $3d^94s$ configurations of $^{76,77}$Se$^{6+}$, these configurations are treated as reference configurations in the subsequent calculations. The $3s$, $3p$, $3d$ subshells in $^{56}\mathrm{Fe}^{8+}$ and $3d$, $4s$ subshells in $^{76,77}$Se$^{6+}$ are defined as valance subshells, and the other inner subshells are core subshells. By allowing restricted single (S) and double (D) substitutions from the reference configurations to active sets with principal quantum numbers up to $n = 7$ and with orbital angular momenta up to $l = 6$, we include the VV correlations for both $^{76,77}$Se$^{6+}$ and $^{56}\mathrm{Fe}^{8+}$, CV correlations of $2s$, $2p$ subshells for $^{56}\mathrm{Fe}^{8+}$, CV correlations of $3s$, $3p$ subshells for $^{76,77}$Se$^{6+}$. The RSCF calculations were followed by RCI calculations, including the Breit interaction and leading QED effects. By using the resulted ASFs, hyperfine and Zeeman interactions were computed by using the HFSZEEMAN program \cite{ANDERSSON2008,LI2020}. 

\begin{table*}\footnotesize
\caption{\label{table} The MCDHF calculations of energies (in eV) of metastable electronic states $3s^23p^53d$ $^3$F$_4$ of $^{56}$Fe$^{8+}$ and $3d^94s$ $^3$D$_3$ of $^{76,77}$Se$^{6+}$ together with their lifetimes (in s) at $B = 0, 3$ and \SI{7}{\tesla}. The nuclear spin for the ground state of $^{76}$Se$^{6+}$ and $^{77}$Se$^{6+}$ are 0 and 1/2, respectively.The $J$ state energies are the mean values of the individual hyperfine states and magnetic sub-states and the lifetimes are averaged from the transition rate calculations assuming a uniform-populated $M_J$ and $M_F$ states. The $J$ state populations are calculated under the collisional-radiative model with an impact electron energy of 170 and \SI{130}{\electronvolt} for $^{56}$Fe$^{8+}$ and $^{76,77}$Se$^{6+}$ ions, respectively. }
\begin{ruledtabular}
\begin{tabular}{rrlcccc}
Ion       & Metastable electronic state  & Energy in eV & & Lifetime in s&  & Population\\

& & & $B=$\SI{0}{\tesla} & $B=$\SI{3}{\tesla} & $B=$\SI{7}{\tesla} &\\
\colrule
$^{56}$Fe$^{8+}$ & $3s^23p^53d$ \quad $^3$F$_4$ & 52.80$^{a}$, 53.05$^{b}$, 53.47$^{c}$,&1110$^{a}$, 578$^{b}$, 493$^{c}$& 1110 & 1110 &19\% \\
& & 52.8$^{d}$, 52.79$^{e}$, 52.79$^{f}$ &970$^{d}$ and 1085$^{e}$&\\
\\
$^{76}$Se$^{6+}$ & $3d^94s$ \quad $^3$D$_3$      & 54.52$^{a}$, 54.52$^{f}$ & 4518	& 563 &	115	& 9\% \\
\\
$^{77}$Se$^{6+}$ & $3d^94s$ \quad $^3$D$_3$ &  54.52$^{a}$ & 865	& 375	& 113	& 9\% \\
\end{tabular}
\end{ruledtabular}
$^{a}$ This work; $^{b}$ S. S. Tayal et al.~\cite{Tayal_2015}; $^{c}$ Au{\v{s}}ra Kynien{\.e}, et al.~\cite{kyniene2019}; $^{d}$ M Hahn et al.\cite{Hahn_2016}; $^{e}$ CHIANTI~\cite{Dere1997}; $^{f}$ NIST~\cite{NIST_ASD}.
\end{table*}

\noindent In Table~\ref{table}, our calculations on the energies of metastable electronic state $3s^23p^53d$ $^3$F$_4$ of $^{56}$Fe$^{8+}$ and $3d^94s$ $^3$D$_3$ of $^{77,76}$Se$^{6+}$ are in good consistency with other theoretical values. For lifetime calculations of $^3$F$_4$ state, it can be noticed that there are several decay channels to the lower states (see Fig.~\ref{figFe}). Among them, the transition rate of $^3$F$_4$ to $^3$P$_2$ determines the lifetime, because other decay channels are even higher-order forbidden transitions, of which the transition rates are much smaller. Further, the transition of the intermediate state $^3$P$_2$ to the ground state is much faster compared with that of $^3$F$_4$ to $^3$P$_2$. With a typical magnetic field in Penning trap, although the wavefunction of $^3$F$_4$ state can mixed with other states like $^3$F$_3$, however, due to its low transition rate the lifetime of $^3$F$_4$ state is not affected. For calculations of $^{77,76}$Se$^{6+}$, this M3 forbidden transition of $3d^94s$ $^3$D$_3$ to ground state can be impacted through hyperfine interaction and external magnetic field because of the strong mixing with shorted-lived $^3$D$_2$ and $^1$D$_2$ states whose decay rates are $4.3\times 10^{4}$ \SI{}{\per\second} and $1.6 \times 10^{5}$ \SI{}{\per\second}, respectively. As a result, for $^{76}$Se$^{6+}$ ion with zero nuclear spin the lifetime of $^3$D$_3$ state reduces from \SI{4518}{\second} to \SI{563}{\second} and \SI{115}{\second} in the external magnetic field from zero to \SI{3}{\tesla} and \SI{7}{\tesla}, respectively, while for $^{77}$Se$^{6+}$ ion with a nuclear spin of 1/2, the lifetime of $^3$D$_3$ state is reduced to \SI{865}{\second} without external magnetic field. In a magnetic field upto \SI{7}{\tesla}, the interaction between the electronic states and the external field becomes much stronger than the interaction with nucleus, which is in the so-called Paschen-Back regime, and the resulting lifetimes of two isotopes are similar. \\
Note that, in a strong magnetic field only $M_J$ and $M_F$ are good quantum numbers instead of $J$ and $F$. Each run of the measurements, a decay from one of the $M$-states is recorded via Penning trap spectrometry. After a number of runs, the lifetime of the $J$ state can be determined by averaging values. The detailed transition rate calculations on $M_J$ and $M_F$ states of $^3$D$_3$ of $^{77,76}$Se$^{6+}$ are listed in the Appendix \ref{apx:tab}:~Tables. As long as decay from different $M$-states is time-resolved, it gives a unique opportunity to identify individual event of transitions from each $M$-states. In table~\ref{table}, the listed lifetime of $^3$D$_3$ state are averaged values assuming a uniform-populated $M_J$ and $M_F$ states. Via a polarized electron beam for instance in a homogeneous strong magnetic field, the population on magnetic sub-states can be non-uniform. According to a population calculation based on a collisional radiative model (CRM) for magnetic sub-states, the relative population percentages for $M_J=\pm3,\pm2,\pm1,0$ in $^3$D$_3$ state fall in a range of $12.2\%$ to $15.9\%$.

\section{\label{sec:level5}Discussion and Implications}
In this section, we will discuss the implementation of energy and lifetime measurements and its future prospect in this proposal manuscript. \\
As a single $^{56}$Fe$^{8+}$ ion is prepared normally in the ground state, it needs to be excited to its metastable electronic state $^3$F$_4$ for measurement. To this end, an electron beam emitted from an FEA (by HeatWave Labs) with a size of \SI{1}{\milli\meter} and current up to mA can be used for EIE. The probability to find an ion staying at this metastable electronic state can be calculated by solving a set of rate equations of excitation, radiation and other process between all the atomic states, which is described by a collisional radiative model. Based on CRM calculations integrated in a flexible atomic code \cite{Gu2008} (for details, see the Appendix  \ref{apx:2}), the metastable electronic states can be produced in sub-seconds with the typical electron density produced by an FEA and the resulting populations are 19\% and 9\% for $^3$F$_4$ of Fe$^{8+}$ and $^3$D$_3$ of Se$^{6+}$, with the impact electron energy of \SI{170}{\electronvolt} and \SI{130}{\electronvolt}, respectively. Although the impact energy is well below the ionization threshold, the collision with electrons can lead to a reduction in the ion's charge through radiative recombination (RR). With a careful control of electron density, this can be avoided since the cross section of RR is 3-4 orders of magnitude lower than EIE. Even if this RR process happened, the electron beam energy can be tuned over the ionization threshold to reproduce the right charge state in short time. Once the electron beam collides with background gas, contamination ions could be created. In this case, a quick axial pulse can be employed to 'kick' other ions out of the trap. To reduce the chance of producing contamination ions, the a short EIE time and a small beam size are favourable.\\
The visibility in the observation of decay from metastable electronic states via the proposed Penning trap spectrometry is mainly limited by the experimental phase jitter of the modified cyclotron motion. This uncertainty in the phase determination consists of cyclotron frequency fluctuations and a technical jitter due to thermal motion and FFT readout. According to our simulations, the technical jitter is about 5 degrees if the signal with a reasonable SNR is tuned, while, the other jitter depends on the magnetic field fluctuation and is proportional to the evolution time. By setting a $2\sigma$ confidence level for the observations of the decay, the corresponding visible phase jump needs to be:

\begin{equation}
\Delta \varphi > 2\sqrt{\delta \varphi_0^2 + \delta \varphi_B^2},
\label{eq:16}
\end{equation}

\noindent here, $\Delta \varphi =360^{\circ} \frac{\Delta M}{M} \nu_+t_\mathrm{evol}$, $\Delta M$ is the equivalent mass difference between the metastable electronic state and the ground state. $\delta \varphi_B = 360^{\circ} \frac{\delta B}{B} \nu _+t_\mathrm{evol}$ and  $\varphi_0$  represents the technical phase jitter. 

\begin{figure}[th!]
\centering
\resizebox{0.5\textwidth}{!}{
\includegraphics{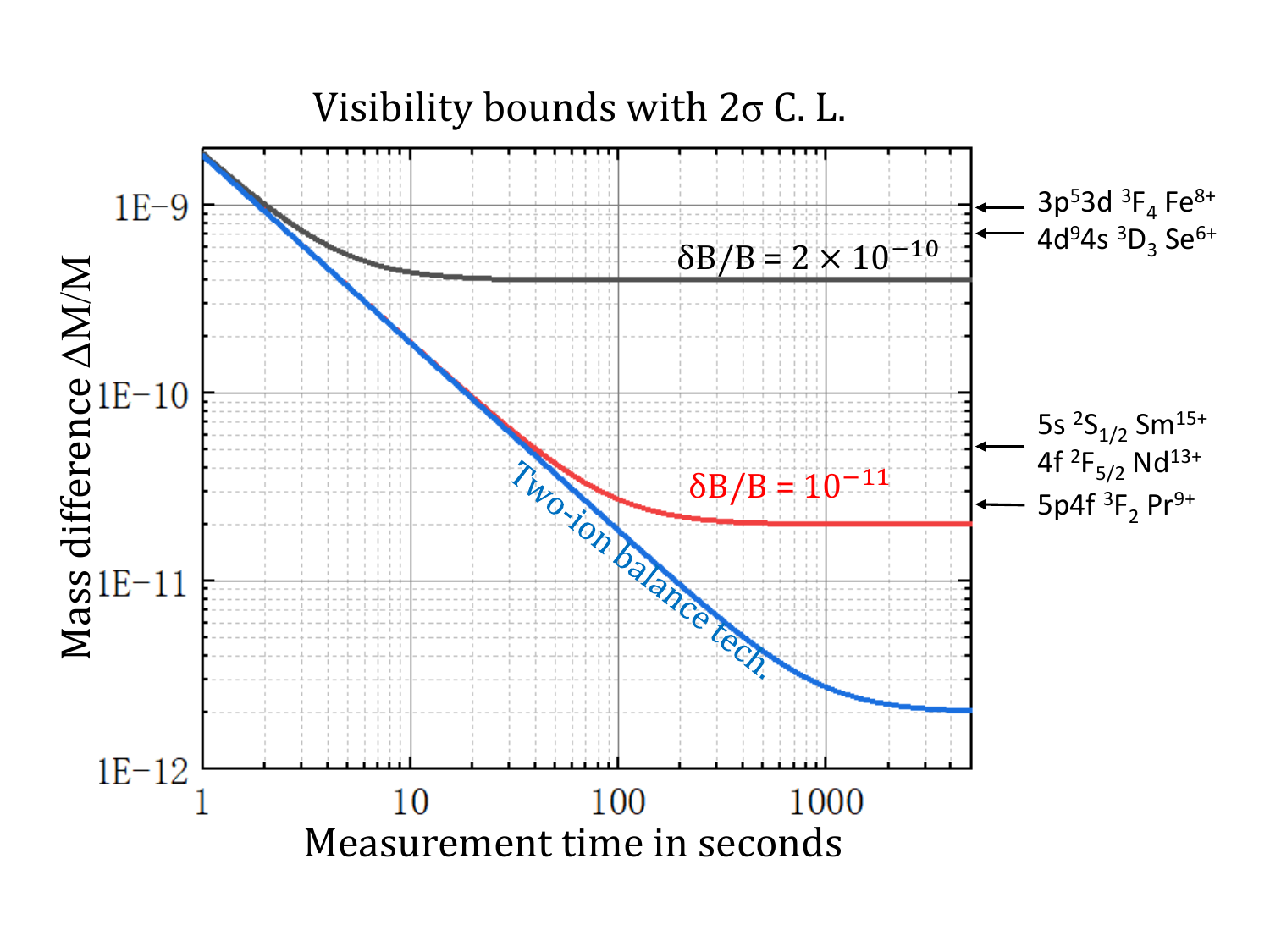}
}
\caption{The $2\sigma$ confidence level (C.L.) of visibility bounds for observations of decay from metastable electronic states with magnetic field fluctuations at $2\times10^{-10}$ (black line) and $10^{-11}$ (red line) levels. With the two-ion balance technique, here assumes that a relative cyclotron phase jitter of $10^{-12}$ can be achieved (blue line). The equivalent mass difference $\Delta M/M$ of long-lived metastable electronic states $5p^53d$ $^3$F$_4$ of Fe$^{8+}$ ($\Delta E = 52.8$ \SI{}{\electronvolt} this work), $4d^94s$ $^3$D$_3$ of Se$^{6+}$ ($\Delta E = 54.5$ \SI{}{\electronvolt} this work), $5s$ $^2S_{1/2}$ of Sm$^{15+}$ ($\Delta E = 7.5$ \SI{}{\electronvolt}) ~\cite{Safronova2014},$4f$ $^2F_{5/2}$ of Nd$^{13+}$ ($\Delta E = 6.9$ \SI{}{\electronvolt})~\cite{Safronova2014}, $5p4f$ $^3F_{2}$ of Pr$^{9+}$ ($\Delta E = 2.8$ \SI{}{\electronvolt})~\cite{Safronova2014} and their ground states are marked.}
\label{fig:7}
\end{figure}

\noindent In Fig.\ref{fig:7}, the visibility bounds with $2\sigma$ confidence level (C.L.) are shown with different magnetic field fluctuations and assuming $\nu_c=$ \SI{15}{\mega\hertz}. For short measurement time, the measurable mass difference is limited by the technical jitter and the free cyclotron frequency. Lower technical jitter and higher cyclotron frequency will extend this visibility bound. Taking into consideration both the ion cooling time and the measurement time, the shortest lifetime that can be measured needs to be longer than \SI{1}{\second}. In principle, to observe decay with even shorter lifetime is possible, but the probability of occurrence of the decay is reduced as $P(t)=\mathrm{exp}(- \frac{t}{\tau})$. For a long measurement time, the measurable mass difference mostly depends on the magnetic field fluctuation. Thus, to extend this limit the stability of the field needs to be improved. The longest lifetime that can be measured by this method is basically limited by the time that the ion can be stored, which depends on the charge exchange cross section and the vacuum in the trap. \\
In Table \ref{table}, the lifetime of metastable electronic state $^3$F$_4$ of $^{56}$Fe$^{8+}$ according to our MCDHF calculations is 1110 s which agrees with CHIANTI database~\cite{Dere1997} but shows large discrepancy to S. S. Tayal et al.~\cite{Tayal_2015} and Kynien{\.e}, et al.~\cite{kyniene2019}. Since this state is barely affected by the typical magnetic field strength in a Penning trap, its lifetime can be directly measured allowing to distinguish between existing predictions.  For metastable electronic state $^3$D$_3$ of  $^{77,76}$Se$^{6+}$, its life time does get impacted by the hyperfine effect and the external magnetic field, but in turn, it provides an opportunity to probe the hyperfine and magnetic quenching.\\
Since both hyperfine and magnetic quenching result in reduction of the lifetime, which means the two effects are in competition. Therefore, if we want to observe the hyperfine quenching, the magnetic quenching effect must be controlled to be small, and vice versa. To this end, the even-even nuclei is a perfect candidate for probing magnetic quenching effect, e.g. $^{76}$Se$^{6+}$ . To measure its lifetime in different fields, changing the main field is one solution but it will take lots of effort and furthermore lowering the magnetic field will always has limit due to the visibility of the phase jump. Another way is to build storage traps on the outer side where the field is low. Once the metastable electronic state is produced and measured in strong-field measurement trap, the ion is then transported to low-field storage trap. This way, the wavefunction of the metastable electronic state turns from a stronger mixing $M$-state to a weaker mixing state whose lifetime is much longer. Then, the electronic state can be determined in MT again after a waiting time. During the waiting time an identical ion can be used in MT to monitor the main magnetic field drift by measuring its cyclotron frequency continuously. In order to probe the hyperfine quenching, isotope nuclei with nuclear spin should be used and strong magnetic field is unfavourable. Since, it is not possible to totally get rid of magnetic field, an alternative solution is still to use storage traps described above and thus the magnetic quenching effect scaling with $B^2$ becomes small while the hyperfine interaction is sizeable. \\
Turn to the future prospect, a direct detection on states such as $5s$ $^2S_{1/2}$ of Sm$^{15+}$, $4f$ $^2F_{5/2}$ of Nd$^{13+}$ and $5p4f$ $^3F_{2}$ of Pr$^{9+}$ (marked in Fig.\ref{fig:7}) for HCI clock transitions  proposed by Safronova et al.~\cite{Safronova2014}  can be expected, after installing dedicated self-shielding coils and pressure regulation system in the liquid helium dewar of superconducting magnet, which has been already mounted in many Penning trap facilities, improves the magnetic field stability to a few $10^{-11}$ level. Further, by combining with a two-ion balance technique \cite{rainville2004,sailer2022} to control two ions with very close charge to mass ratio rotating in a common magnetron orbit, we could measure relative phases of cyclotron motion of the two ions to balance the inherent fluctuations of the magnetic field on the individual ions. Based on this two-ion balance system, an optical laser can be employed to excite one of the ion. Once the ion is pumped from one state to the other, the mass variance can be observed by applying a simultaneous modified cyclotron phase measurement, which allows for a precision towards $10^{-12}$ level and paves the way to distinguish most long-lived metastable electronic states to search for suitable HCI clock transitions.

\section{\label{sec:level6}Conclusion}
In summary, we present an experimental access to observe the decay of long-lived metastable electronic states of highly charged ions using a Penning trap. To measure the lifetimes of long-lived metastable electronic states, which is very challenging by any other conventional techniques, becomes in principle feasible through the proposed techniques based on a single-ion mass spectrometry. The method's implementation and benefits are described in detail. A dedicated simulation study has been conducted to validate the method's effectiveness and discuss expected results in a realistic scenario. For the theoretical studies, we calculated the energy levels and transition rates of $3p^53d$ $^3$F$_4$ state of $^{56}$Fe$^{8+}$ and $3d^94s$ $^3$D$_3$ state of $^{77,76}$Se$^{6+}$ which are suitable candidates for testing the proposed technique. Furthermore, the expected outcomes of measuring their lifetimes to distinguish the existed theories and to probe hyperfine and magnetic field quenching effects are previewed. This method will be potentially extended to use in any precision Penning trap mass spectrometry to detect metastable electronic states with a broad range of energy and lifetimes for fundamental research purposes. 

\section{Acknowledgments}
The author thanks Prof. Klaus Blaum, Dr. Sven Sturm, Dr. Fabian Hei\ss er, Dr. Wenxian Li and Dr. Xiangjin Kong for kind discussion. This work was supported by the National Key R{\&}D Program of China under Grant No. 2022YFA1602504 and No. 2022YFA1602303, the National Natural Science Foundation of China under Grant No. 12204110, No. 12074081 and No. 12104095, sponsored by Shanghai Pujiang Program under Grant No. 22PJ1401100 and Max-Planck Partner Group Project.

\appendix

\section{\label{apx:tab}Tables}
Here, we listed the calculated transition rates in $M_J$ and $M_F$ states for $^3$D$_3$ of $^{77,76}$Se$^{6+}$ ions with magnetic field of 0, 3 and \SI{7}{\tesla} in tables~\ref{tabSe76mit}-\ref{tabSe77mit5}.

\begin{table}\footnotesize
\caption{\label{tabSe76mit} Transition rates (s$^{-1}$) of $3d^94s\ ^3D_3$  (in $M_J'$ states) $\rightarrow 3d^{10}\ ^1S_0$ (in $M_J$ states) in presence of magnetic field $B = \SI{3}{\tesla}$ and $B = \SI{7}{\tesla}$ for $^{76}\rm{Se}^{6+}$.}
\begin{ruledtabular}
\begin{tabular}{rllllll}
 $M_J'$ & $M_J$  & \multicolumn{2}{c}{Rates}\\
\colrule
& & $B=3T$ & $B=7T$\\
  3 & 0 & 2.21E-04 & 2.21E-04\\
  2 & 0 & 1.78E-03 & 8.71E-03\\
  1 & 0 & 2.71E-03 & 1.38E-02\\
  0 & 0 & 3.02E-03 & 1.55E-02\\
  1 & 0 & 2.71E-03 & 1.38E-02\\
  2 & 0 & 1.78E-03 & 8.68E-03\\
  3 & 0 & 2.21E-04 & 2.21E-04\\
\end{tabular}
\end{ruledtabular}
\end{table}

\begin{table}\footnotesize
\caption{\label{tabSe77hfs} Transition rates (s$^{-1}$) of $3d^94s\ ^3D_3$ (in $F'$ states) $\rightarrow 3d^{10}\ ^1S_0$ in absence of magnetic field $B = \SI{0}{\tesla}$ for $^{77}\rm{Se}^{6+}$.}
\begin{ruledtabular}
\begin{tabular}{llll}
 $F'$  & Rates\\
\colrule
 7/2 & 2.21E-04\\
 5/2 & 2.09E-03\\
\end{tabular}
\end{ruledtabular}
\end{table}

\begin{table}\footnotesize
\caption{\label{tabSe77mit7} Transition rates (s$^{-1}$) of $3d^94s\ ^3D_3\ F=7/2$ (in $M_F'$ states) $\rightarrow 3d^{10}\ ^1S_0\ F=1/2$ (in $M_F$ states) in presence of magnetic field $B = \SI{3}{\tesla}$ and $B = \SI{7}{\tesla}$ for $^{77}\rm{Se}^{6+}$.}
\begin{ruledtabular}
\begin{tabular}{rrlllll}
 $M_F'$ & $M_F$  & \multicolumn{2}{c}{Rates}\\
\colrule
& & $B=3T$ & $B=7T$\\
  7/2 &  1/2 & 2.21E-04 & 2.21E-04\\
  5/2 &  1/2 & 8.80E-04 & 1.09E-03\\
  5/2 & -1/2 & 2.14E-04 & 2.20E-04\\
  3/2 &  1/2 & 3.06E-05 & 1.75E-04\\
  3/2 & -1/2 & 2.18E-04 & 2.63E-04\\
  1/2 &  1/2 & 9.57E-05 & 1.57E-03\\
  1/2 & -1/2 & 2.72E-04 & 1.26E-03\\
 -1/2 &  1/2 & 2.29E-04 & 2.29E-03\\
 -1/2 & -1/2 & 5.48E-04 & 3.65E-03\\
 -3/2 &  1/2 & 1.74E-04 & 1.49E-03\\
 -3/2 & -1/2 & 8.96E-04 & 6.17E-03\\
 -5/2 &  1/2 & 2.38E-06 & 5.82E-07\\
 -5/2 & -1/2 & 9.56E-04 & 6.25E-03\\
 -7/2 & -1/2 & 2.21E-04 & 2.21E-04\\
\end{tabular}
\end{ruledtabular}
\end{table}

\begin{table}\footnotesize
\caption{\label{tabSe77mit5} Transition rates (s$^{-1}$) of $3d^94s\ ^3D_3\ F=5/2$ (in $M_F'$ states) $\rightarrow 3d^{10}\ ^1S_0\ F=1/2$ (in $M_F$ states) in presence of magnetic field $B = \SI{3}{\tesla}$ and $B = \SI{7}{\tesla}$ for ion $^{77}\rm{Se}^{6+}$.}
\begin{ruledtabular}
\begin{tabular}{rrllll
l}
 $M_F'$ & $M_F$   & \multicolumn{2}{c}{Rates}\\
\colrule
& & $B=3T$ & $B=7T$\\
  5/2 &  1/2 & 4.06E-03 & 1.25E-02\\
  5/2 & -1/2 & 6.89E-06 & 9.32E-07\\
  3/2 &  1/2 & 5.93E-03 & 1.96E-02\\
  3/2 & -1/2 & 3.99E-05 & 1.99E-05\\
  1/2 &  1/2 & 6.34E-03 & 2.18E-02\\
  1/2 & -1/2 & 1.52E-04 & 9.99E-05\\
 -1/2 &  1/2 & 5.48E-03 & 1.93E-02\\
 -1/2 & -1/2 & 3.90E-04 & 2.97E-04\\
 -3/2 &  1/2 & 3.43E-03 & 1.20E-02\\
 -3/2 & -1/2 & 7.93E-04 & 6.74E-04\\
 -5/2 &  1/2 & 2.19E-04 & 2.21E-04\\
 -5/2 & -1/2 & 1.40E-03 & 1.30E-03\\
\end{tabular}
\end{ruledtabular}
\end{table}

\section{\label{apx:2}CRM calculation}
To calculate the probability of producing metastable electronic states by EIE in a specific atomic level system like $^{56}$Fe$^{8+}$, a sophisticated method known as collisional-radiative model can be used \cite{HARTGERS200}. In this model, atomic processes such as spontaneous radiation, electron impact excitation and de-excitation, recombination, ionization and charged exchange are taken to form a set of differential rate equations expressing the population and de-population of atomic levels. In the case of $^{56}$Fe$^{8+}$, the energy of metastable electronic state $^3$F$_4$ is \SI{52.79}{\electronvolt} which is much lower than the ionization threshold and thus by tuning the electron impact energy below this threshold the ionization process can be excluded. The recombination process can reduce the charge state. However, the cross section of the radiative recombination is 3-4 order of magnitude lower than the EIE process. With a pulsed electron beam, it is possible to avoid an incident of radiative recombination. In contrast to radiative recombination, the resonant dielectric recombination has a much large cross section comparable to EIE. Nevertheless, by tuning the electron beam energy off the resonance, dielectric recombination can also be avoided. In a measurement time of a range from hundreds of seconds to weeks, the charged exchange can hardly happen due to the good vacuum under the cryogenic environment. In the end, the differential rate equations can be expressed as: 

\begin{equation}
\begin{aligned}
\frac{dN_i}{dt}&=\sum_{j>i}A_{j\rightarrow i}^rN_j+\sum_{j<i}C_{j\rightarrow i}^eN_jn_e+\sum_{j>i}C_{j\rightarrow i}^dN_jn_e\\
&-\sum_{j<i}A_{i\rightarrow j}^rN_i-\sum_{j>i}C_{i\rightarrow j}^eN_in_e-\sum_{j<i}C_{i\rightarrow j}^dN_in_e,
\end{aligned}
\label{eq:17}
\end{equation}

\noindent where, the subscript $i$,$j$ represent the initial stat and final state and $N_i$ denotes the population of state $i$. $A_{j\rightarrow i}^r$, $C_{j\rightarrow i}^e$ and $C_{j\rightarrow i}^d$ represent the radiative decay rate, the cross sections of electron impact excitation and electron impact de-excitation, respectively. $n_e$ denotes the electron density. According to the equilibrium condition $\frac{dN_i}{dt}=0$ and normalized condition $\sum N_i =1$ the population of each state can be calculated. \\

\bibliographystyle{unsrtnat}
\bibliography{Tu}  

\end{document}